\documentclass[prstab,showkeys,ajp,preprintnumbers,amsmath,amssymb,runinaddress,preprint]{revtex4}
\usepackage[dvips]{graphicx}
\usepackage{dcolumn}
\usepackage{bm}

\begin{document}
\title{Reduction of the Casimir force using  aerogels}

 \author{R. Esquivel-Sirvent}

\affiliation{Instituto de F\'{i}sica, Universidad Nacional Aut\'{o}noma de M\'{e}xico \\
Apdo. Postal 20-364, Ciudad Universitaria, D. F. 01000, M\'{e}xico.}

 \date{\today}

\begin{abstract}
  By using silicon oxide based aerogels we show numerically that the Casimir force can be reduced several orders of magnitude,  making its effect negligible in nanodevices. This decrease in the Casimir force is also present even when the aerogels are deposited on metallic substrates. To calculate the Casimir force we model the  dielectric function of silicon oxide aerogels using an effective medium dielectric function such as the Clausius-Mossotti approximation.  
The results show that both the porosity of the aerogel and its thickness can be use as control parameters to reduce the magnitude of the Casimir force.     
\end{abstract}


\maketitle

\section{Introduction}

The last decade has seen a renewed interest in the study of the Casimir force \cite{casimir}. Casimir predicted  that two neutral parallel plates will attract each other due to a force arising from quantum vacuum fluctuations.  
The interest in this force arises from the precise measurements  between metallic \cite{lamoreaux,iannuzzi,decca,mohideen,bressi} and dielectric systems \cite{mohideen2,mohideen3} that are now possible. 
The measurements have confirmed the validity of Casimir's theory and Lifshitz formulation \cite{lifshitz}, that takes into account the dielectric properties of the plates.  The Lifshitz theory gives the Casimir force or the Van der Waals force in  the retarded and non-retarded limit between extended bodies.  Usually, in the literature some authors use the term Van der Waals force in  the non-retarded limit and the term Casimir force  in the retarded limit \cite{bordag}.  

Recent research in Casimir forces is focused in having a
 precise comparison between theory and experiment. To achieve this, several physical characteristics of the plates have to be known such as 
conductivity \cite{lamoreaux2,bostrom,genet}, roughness \cite{palas1,palas2} and finite temperature effects and its  controversies due to the apparent violation of Nerst heat theorem \cite{schaden,chen,brevik07,mostepanenko07}.  

The Casimir force has also been considered as a possible actuating force in micro and nano electromechanical devices (MEMS and NEMS).
In 1995 Serry {\it et al.} were the first to consider the role that Casimir forces could have in MEME by analyzing the dynamics of a micro system actuated only by Casimir forces \cite{serry1}. Later, this system  was realized experimentally by Chan \cite{capasso1} who also  able to excite a micro torsional balance in its non linear mode using only the Casimir force \cite{ capasso2}.  A recent review of Casimir forces in nanomechanics an nanotechnology can be found in Reference \cite{capasso3}.
Later, the deflection of micro membrane strips under a purely Casimir load was also studied by Serry \cite{serry2}.  These works suggested that 
another consequence of the Casimir force in micro and nano devices, was that it   could limit the operation of  modern MEMS and NEMS by causing a pull-in or snap down instability similar to what happens to electrostaticaly actuated devices \cite{pelesko,pelesko02}.  The stability of MEMS in the presence of Casimir has also been studied by several authors 
 \cite{zhao,guo04,zhao07,esquivelapl,esquivelnjp} as well as the pull-in dynamics when both electrostatic and Casimir forces are present \cite{batra}.
However, as shown by Delrio $et al. $ \cite{delrio}, the  Casimir force  is relevant in the pull-in dynamics  in the non retarded limit (or Van der Waals regime). In the case of metals, the nonretarded limit corresponds to separations smaller than the plasma wavelength of the material.  
  
      Unlike the electrostatic case, the Casimir force can not be 'turned-off' and is always present. Its influence can be reduced in several ways. For example, using  thin conducting films will reduce the Casimir force \cite{lisanti,njp07} or  using low dielectric function materials such as $Si$ \cite{mrs,irina,mohideen06} . It has also been suggested that it is possible to change the sign of the force from attractive to repulsive using high magnetic susceptibility materials \cite{klich,comment}.

     In this work we show that the Casimir force can be reduced drastically using silicon based aerogels.  In particular, we discuss the use of  $SiO_2$ aerogels and how the dielectric function of the aerogels rescales the Casimir force.  
\section{Formalism}
The calculations are done in the Lifshitz approach at zero temperature and assuming only frequency dependent dielectric functions.  

The Casimir force per unit area between two arbitrary flat surfaces labeled $ j=1,2$ separated a distance $L$ 
is calculated using the Lifshitz formula \cite{lifshitz} 
 \begin{equation}
       \label{lifshitz}
       F=\frac{\hbar c }{2 \pi^{2}}\int_{0}^{\infty}Q dQ\int_{q>0}dk\frac{k^{2}}{q}(G^{s}+G^{p}),
\end{equation}
where $G_s= (r_{1s}^{-1} r_{2s}^{-1} \exp{(-2  k L )}-1)^{-1}$ and $G_p=(r_{1s}^{-1} r_{2s}^{-1} \exp{(-2  k L
)}-1)^{-1}$. In these expressions, the factors
  $r_{p,s}$  are the reflectivities for either $p$ or $s$ polarized light , $Q$ is the wavevector component along the
plates, $q=\omega/c$ and $k=\sqrt{q^2+Q^2}$. The above expression is evaluated along the imaginary frequency axis $i \omega$. 

 When the reflectivities are replaced by the Fresnel coefficients, the
original Lifshitz formula is recovered. For perfect conductors,   $|r_{p}|=|r_{s}|=1$  and Eq. (\ref{lifshitz}) yields Casimir's known expression for the force in the ideal case, 
\begin{equation}
\label{ideal}
F_c= -\frac{\hbar c \pi^2}{240 L^4}.
\end{equation}
  
 In general, the problem of calculating the Casimir force between two surfaces   
 reduces to finding the reflection coefficients given dielectric function of the materials.  
   For a slab of thicknesss $D$ and dielectric function $\epsilon_1$ on top of a half space of a material characterized by  a dielectric function $\epsilon_2$ the reflection amplitudes are
\begin{equation}
r_{p,s}=\frac{r^{01}_{ps}+r^{12}_{ps}e^{-2\delta}}{1+r^{01}_{ps}r^{12}_{ps}e^{-2\delta}}
\label{reflect}
\end{equation}
where $r^{i,j}_{p,s}$  are the Fresnel coefficients between material $i$ and $j$ where the subindex $0$ stands for vacuum.  The optical length is defined as \cite{born}
\begin{equation}
\delta=\frac{D}{c}\sqrt{\omega^2(\epsilon_1(i \omega)-1)+c^2 k^2}. 
\label{phase}
\end{equation} 
These definitions of the reflection amplitude and optical length are in the rotated space $i\omega$. This is, we change,  $\epsilon(\omega)\rightarrow \epsilon(i \omega)$.  

\section{Results}
In this work we consider the use of aerogels to reduce the magnitude of the Casimir force. Aerogels  are  highly porous materials fabricated typically from $SiO_2$ using sol-gel techniques. The porosity is as high as $95\%$ and they are promising materials in the semiconductor industry \cite{maex} due to their low thermal conductivity, low density, good mechanical strength to compressional loads and their luminescent properties make them suitable in high energy physics detectors \cite{sumiyoshi}.  The property we are interested in is the dielectric function. Aerogels are the solids with the lowest index of refraction  in the visible range   $n\sim 1.05$ depending on the porosity.  The dielectric behavior of aerogels are dominated by the air voids rather than by the solid part \cite{hrubesh} and the index of refraction varies linearly with porosity, as has been shown in several measurements in limited frequency ranges \cite{geis,aaa,hrubesh}.   To calculate the Casimir force we need the dielectric response of the aerogel in a broad frequency range. This is done using
an effective medium formalism such as a Clausius-Mossotti approximation.  This approximation is adequate given the high porosity of the aerogels . Other formalisms, such as  Looyenga's approach \cite{looyenga}  have also been used to describe the effective optical properties of porous media at lower porosities.
   Taking the dielectric function of air equal to one for all frequencies, the dielectric function of $\epsilon(\omega)$ of the aerogel is obtained from 
 
 \begin{equation}
 \label{effective}
 \frac{\epsilon-\epsilon_{SiO2}}{\epsilon+2\epsilon_{SiO2}}=\phi \frac{1-\epsilon_{SiO2}}{1+2 \epsilon_{SiO2}},
 \end{equation}  
 where $\phi$ is the porosity of the sample.  The dielectric function of the $SiO_2$ can be obtained from tabulated data \cite{palik}. The tabulated data expands a wide frequency range from 0.00218 eV to 2000 eV being enough to calculate the  Casimir force for separations between $0.1-1.$ $\mu m$. 
  The dielectric function in the imaginary frequency domain is obtained from 
  \begin{equation}
  \epsilon_{SiO2}(i \omega)= 1+\frac{2}{\pi}\int_{0}^{\infty}\frac{w' \epsilon"(w ')}{w'^2+\omega^2}dw',
  \label{epsilica}
  \end{equation}
  where $\epsilon"(w ')$ is the imaginary part of the dielectric function that is obtained from the tabulated data. 
  With the dielectric function of the silicon oxide and Eq.(\ref{effective}) the reflection coefficients and hence the Casimir force can be calculated. In this work we calculate the reduction factor define as $F_r=F/F_c$ where $F$ is given by Eq.(\ref{lifshitz}) and $F_c$ is the ideal Casimir force (Eq.(\ref{ideal})).

Let us consider first the Casimir force between two aerogel layers of thickness $D$ deposited on top of $Au$ substrates. The substrates are assumed as half-spaces and the dielectric function of $Au$ is modeled using a Drude model with a plasma frequency of $\omega_p=9$ $eV$ and a damping constant of $\gamma_D=0.035 eV$.  Depending on the values of the Drude parameters and the tabulated data used, variations of few percents on the calculated force are expected as discussed recently by Piroshenko \cite{irinaau}.   Although the use of the Drude model at finite temperature has been questioned due to the apparent violation of Nerst heat theorem, we will not elaborate in this issue since we are interested in the effect of the aerogel on the Casimir force \cite{mostepanenko07}.

 The reduction factor as a function of separation for different thicknesses of the aerogel layers is shown in Fig. 1.  As a reference we have also plotted the reduction factor for a $Si$ layer $500$ $nm$ thick since this is a common dielectric that has been studied in connection with the Casimir force.  For small separations, the reduction factor is of the order or $10^{-2}$. This is  close to  one tenth of the expected value of the $Si$ layer.  The thicker the value of the aerogel layer the smaller and flatter the reduction factor becomes in a wider range of the separation.   This behavior can be understood considering that for a given frequency and value of the normal wave vector, the optical length increases as $D$  increases. This, in turn, implies that the exponential term in eq.(\ref{reflect}) becomes smaller for thicker slabs and the dominant contribution to the reflection comes from $r_{01}^{s,p}$. For thinner slabs the term containing the exponential has a bigger contribution from the aerogel-Au reflection amplitude $r_{1,2}$ hence, the overall reflection amplitude is bigger and  the force increases.  In Figure 1 and subsequent plots the separation between the slabs ranges from $100$ $nm$ to $2$ $\mu m$ that is a reasonable experimental range to probe the Casimir force. Also we assume that the porosity of the aerogel is $\phi=90\%$. 
 
The measurements of the Casimir force is commonly made between a $Au$ \cite{aucomment} and a second surface that can be a metal or semiconductor.  A more realistic configuration in our calculations is to consider the force between a $Au$ half space and a aerogel supported also on a $Au$ substrate as shown in the inset of Figure 2. Also, in this figure the reduction factor is presented again for various values of the aerogel thickness.  The force is slightly larger than in Figure 1, but even in this case there is a significant reduction.  From both Figure (1) and (2) we see that the aerogel layer acts as a buffer than effectively increase the distance between the $Au$ slabs, thus decreasing the force.

The role of the $Au$ substrate is also  important and if it is remove the reduction factor shows a different behavior as shown in Figure 3.  In this case we plot the reduction factor as a function of plate separation $L$ for slabs of thickness $D$. The reduction factor decreases as the separation increases in all cases.  Indeed the force can be reduced several orders of magnitude for large separations depending on the aerogel slab thickness.  For all practical purposes, aerogels will suppress the Casimir force below the range of detectability of current experiments even at short separations.

Finally let us consider the role of porosity.  In Figure 4 we plot the  reduction factor for two unsupported aerogel slabs of thickness $D=500$$nm$  for different porosities. As expected as the porosity increases, the dielectric function is lower and the Casimir force decreases. The inset shows the value of the reduction factor at  a fixed separation of $L=1$$\mu m$, again for different porosities.  Notice that the vertical axis of the inset has to be multiplied by $10^{-5}$.  This shows that the porosity can be an effective control parameter to modulate the value of the Casimir force.  Of particular importance will be the control of pull-in instabilities in micro and nano devices \cite{esquivelnjp}.  

\section{Conclusions}
In conclusion,  in this work we presented the first numerical study of the use of  aerogels to control the Casimir force by reducing its value significantly.  There are two control parameters that can be changed to modify the force. One is the thickness slab and the second one is the porosity of the aerogel. Also, the role of a metallic substrate is evaluated.  Even when the aerogels have a Au backing the force can be up to orders of magnitude smaller than for perfect metals. As compared to bulk $Si$, aerogels can decrease the Casimir force in several orders of magnitude. 
 Other options that can be consider to reduce the Casimir force is to use porous semiconductor composites such as $GaAs$ or porous $Si$ \cite{kochergin}.

  \acknowledgements{Partial support from CONACyT-Mexico Grant 44306 and DGAPA-UNAM IN-101605. }

\newpage

 \begin{figure}
 \includegraphics[width=8.5cm]{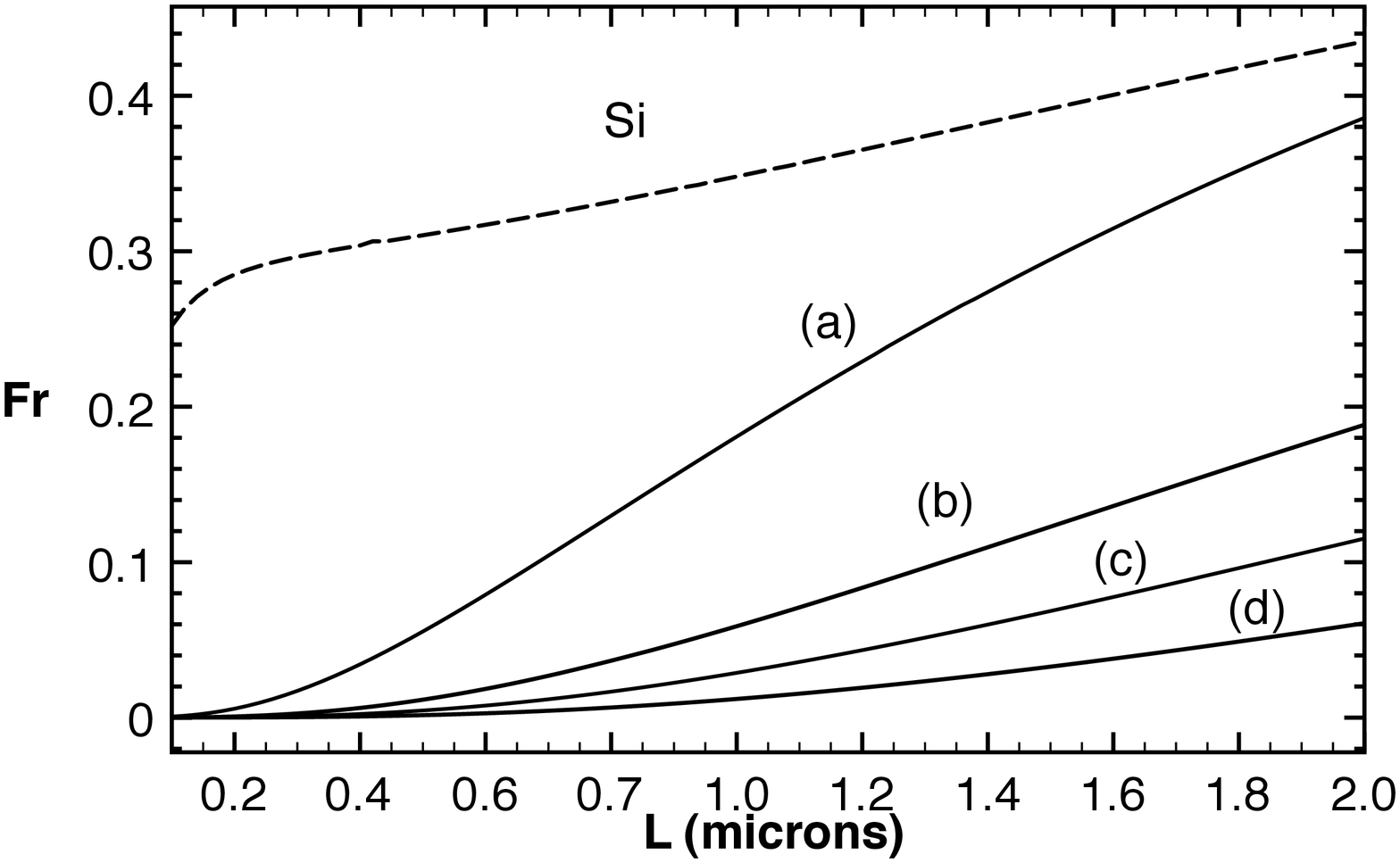}
 \caption{Reduction factor between slabs of aerogel supported on $Au$ substrates, for different aerogel thicknesses. As a reference the reduction factor for a $500$$nm$ thick $Si$ slab is presented.  The thicker the aerogel slab, the smaller the force in a wider range of separations as explained in the text. The calculations are for an aerogel with a $90\%$ porosity. The thicknesses of the aerogel used in the different curves are a) $D=250 nm$, b) $D=500 nm$, c) $D=700 nm$, d)$D=1000 nm$.} 
 \end{figure}
 
 \begin{figure}
 \includegraphics[width=8.5cm]{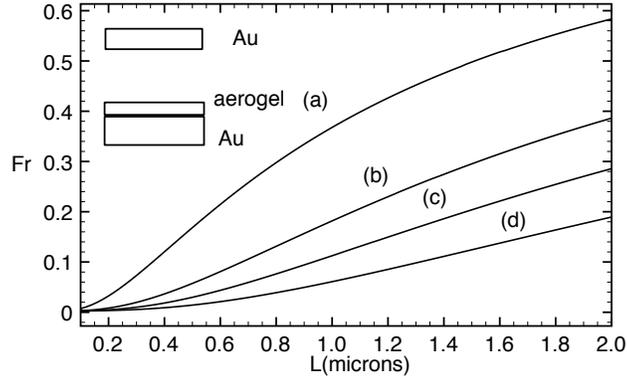}
\caption{The reduction factor is shown for the system depicted in the inset, for different thicknesses of the aerogel layer.   This configuration resembles closer an experimental set up. The same thicknesses $D$ of the aerogel as in Fig.1 were used. Even in this case a significant reduction in the force is obtained. The aerogel slab acts as a buffer that increases the optical distance between the metallic components.}
 \end{figure}
 
 \begin{figure}
 \includegraphics[width=8.5cm]{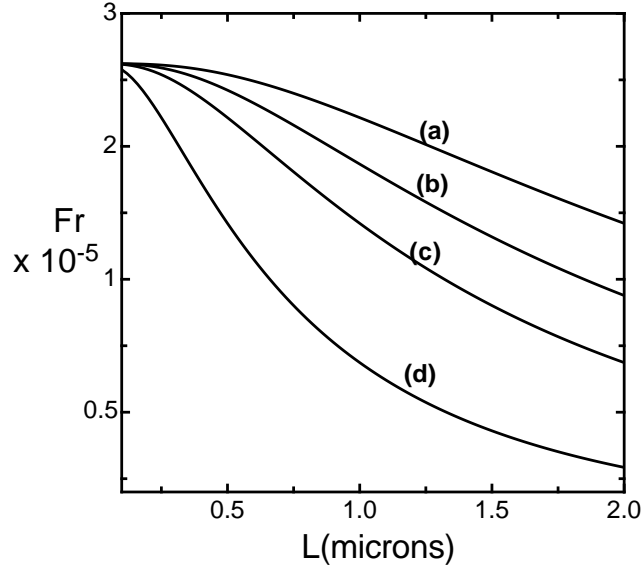}
 \caption{Reduction factor for unsupported aerogel slabs of varying thickness $D$ (no substrate).  For short and large separations a significant reduction of the force can be obtained.  It is important to notice that in this case the thicker the slab the reduction factor increases, contrary to what happens in Fig1. and Fig.2 when a metallic substrate is present. The different curves are for a)$ D=1000nm$, b) $D=700 nm$, c) $D=500 nm$, d) $D=250 nm$. }
 \end{figure}
 \begin{figure}
 \includegraphics[width=8.5cm]{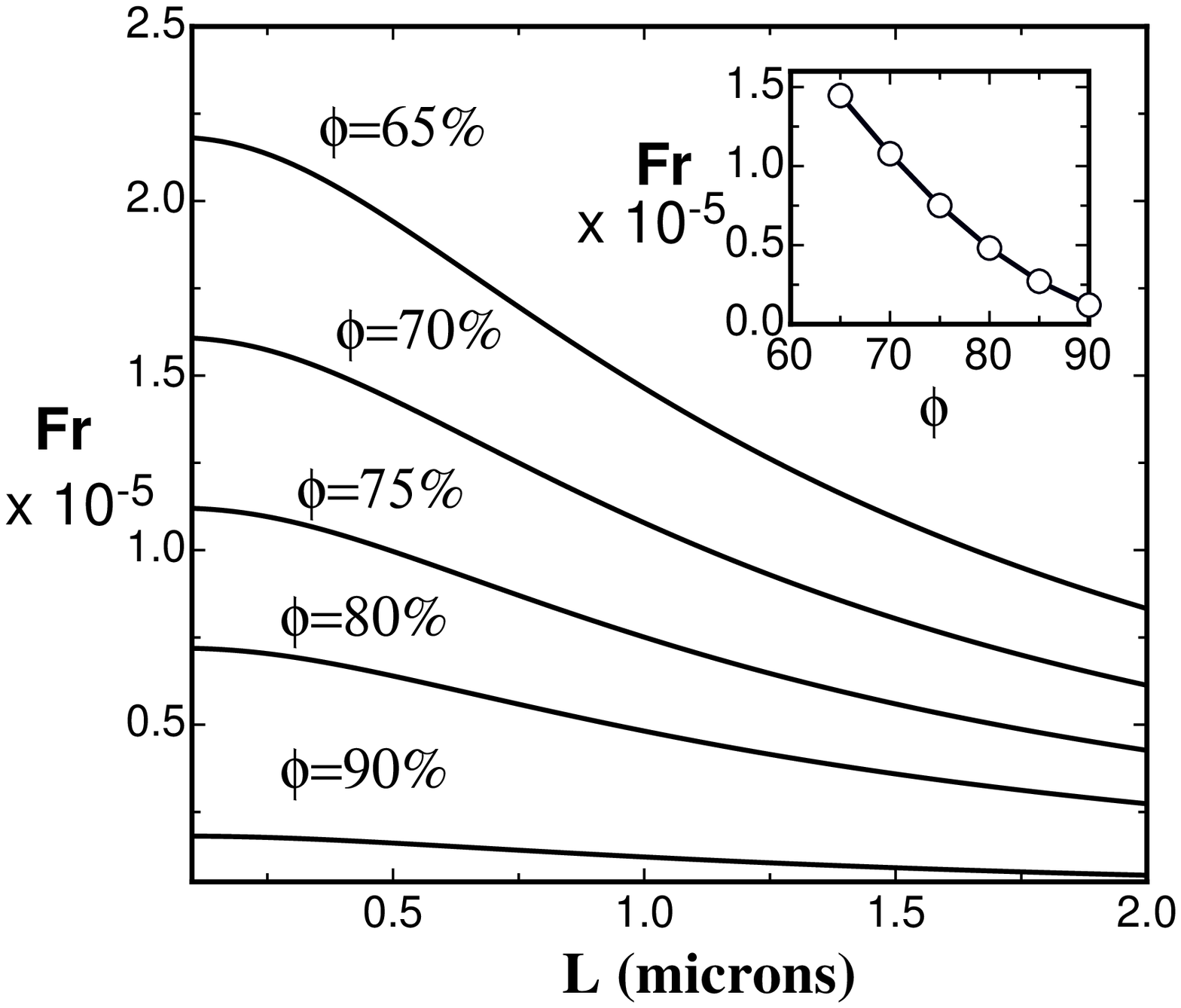}
  \caption{Reduction factor as a function of separation between unsupported slabs of thickness $D=500$$nm$ for different porosities.  As the porosity increases the reduction factor scales down. This is seen in the inset where the maximum value of the reduction factor was plotted as a function of porosity for a fixed separation $L=1 \mu m$.}  \end{figure}
  

\newpage
\begin{references}
\bibitem{casimir} G. Casimir, Proc. K. Ned. Aka H. B. d. Wet. {\bf 51}, 793
(1948).
\bibitem{lamoreaux} S. K. Lamoreaux, Phys. Rev. Lett. {\bf 78}, 5 (1997); {\bf %
81}, 5475 (1998).
 \bibitem{iannuzzi} D. Iannuzzi, M. Lisanti, and F. Capasso, Proc. National
Acad. Sci. USA, {\bf 101}, 4019 (2004).
 \bibitem{decca}R. S. Decca, D. L\'{o}pez, E. Fischbach, and D. E. Krause,
Phys. Rev. Lett. {\bf 91}, 050402 (2003).
 \bibitem{mohideen} U. Mohideen and A. Roy, Phys. Rev. Lett. {\bf 81}, 4549 (1998).
 \bibitem{bressi}G. Bressi, G. Carugno, R. Onofrio, and G. Ruoso, Phys.
Rev. Lett. {\bf 88}, 041804 (2002).
\bibitem{mohideen2} F. Chen, U. Mohideen, G. L. Klimchitskaya, and V. M. Mostepanenko, Phys. Rev. A, {\bf 74}, 022103 (2006). 
\bibitem{mohideen3} F. Chen, G. L. Klimchitskaya, V. M. Mostepanenko, and U. Mohideen, Phys. Rev. Lett. {\bf 97}, 170402 (2006). 
  \bibitem{lifshitz}  E. M. Lifshitz, Sov. Phys. JETP {\bf 2}, 73 (1956).
  \bibitem{bordag} M. Bordag, U. Mohideen and V.M. Mostepanenko, Phys. Rep. {\bf 353}, 1 (2001).
\bibitem{lamoreaux2} S. K. Lamoreaux, Phys. Rev. A {\bf 59} R3149 (1999).
\bibitem{bostrom} M. Bostrom and B. E. Sernelius, Phys. Rev. A {\bf 61} 046101 (2000). 
\bibitem{genet} C. Genet, A. Lambrecht and S. Reynaud, Phys. Rev. A {\bf 62}, 012110 (2000). 
\bibitem{palas1} G. Palasantzas, J. Appl. Phys. {\bf 97}, 126104 (2005).
\bibitem{palas2}  G. Palasantzas, J. Appl. Phys. {\bf 98}, 034505 (2005).
\bibitem{schaden} M. Schaden and L. Spruch, Phys. Rev. A {\bf 65}
034101 (2002).
\bibitem{chen} F. Chen, G. I. Klimchitskaya, U. Mohideen and V. M. Mostepanenko, Phys. Rev. Lett. {\bf 90} 160404 (2003).
\bibitem{brevik07} J. S. Hoye, I. Brevik, S. A. Ellingsen and J. B. Aarseth,  arXiv quant-ph/0703174 (2007). 
\bibitem{mostepanenko07}  V. M. Mostepanenko, V. B. Bezerra, R. S. Decca, B. Geyer, E. Fischbach, G. L. Klimchitskaya, D. E. Krause, D. Lpez and C Romero, J. Phys. A: Math. Gen. {\bf 39}, 6589 (2006).
\bibitem{serry1} F.M. Serry, D. Walliser and G. J. Maclay, IEEE J. Microelectromechanical Systems {\bf 4}, 193 (1995).
\bibitem{capasso1} H. B. Chan, V. A. Aksyuk, R. N. Kleiman, D. J. Bishop, and
F. Capasso, Science {\bf 291}, 1941 (2001).
\bibitem{capasso2} H. B. Chan , V. A. Aksyuk, R. N. Kleiman, D. J. Bishop, and F. Capasso,  Phys. Rev. Lett. {\bf
87}, 211801 (2001).
\bibitem{capasso3} F. Capasso, J. N. Munday, D. Iannuzzi and H. B. Chan, IEEE Journal of Selected Topics in Quantum Electronics, {\bf 13} 400 (2007).
\bibitem{serry2}   F.M. Serry, D. Walliser and G. J. Maclay,  J. Appl. Phys. {\bf 84}, pp. 2501-2506,
(1998).
\bibitem{pelesko} J. A. Pelesko and D. H. Bernstein, {\it Modeling MEMS and NEMS} , (Chapman and Hall/CRC, Boca Raton, Florida, 2003).
\bibitem{pelesko02} J. A. Pelesko, SIAM J. Appl. Math., {\bf 62}, pp. 888-908 (2002).
 \bibitem{zhao}Y. P. Zhao, L. S. Wang and T. X. Yu, J. Adhesion Sci.
Technol., {\bf 17}, 519 (2003).
\bibitem{guo04} J. G. Guo and Y. P. Zhao, J. Microelectr. Sys. {\bf 13}, 1 (2004).
\bibitem{zhao07} Wen-Hui Lin and Ya-Pu Zhao,  J. Phys. D: Appl. Phys. {\bf 40} 1649 (2007). 
  \bibitem{esquivelapl} J. Barcenas. L. Reyes and R. Esquivel-Sirvent, Appl. Phys. Lett. {\bf 87}, 263106 (2005).
  \bibitem{esquivelnjp} L. Reyes, J. Barcenas y R. Esquivel-Sirvent, New J. Phys. {\bf 8}, 241 (2006). 
\bibitem{batra} R. C. Batra, M. Porfiri and D. Spinello, Eur. Phys. Lett. {\bf 77}, 20010 (2007).
\bibitem{delrio} Delrio, F. W., M. P. DeBoer, J. A. Knapp, E. D. Reedy, P. J. Clews and M. L. Dunn, Naturematerials {\bf 4}, 629 (2005).
  \bibitem{irina} A. Lambrecht, I. Pirozhenko, L. Duraffourg and Ph. Andreucci, Eur. Phys. Lett. {\bf 77} 44006 (2007). 
\bibitem{mrs} R. Esquivel-Sirvent and C. Villarreal, Mat. Res. Soc. Proc. {\bf 741}, j5.21.1 (2003). 
\bibitem{lisanti} M. Lisanti, D. Iannuzzi and F. Capasso, Proc. Nat. Acad. Sci. {\bf102}, 11989 (2005).
\bibitem{njp07}  S de Man and D. Iannuzzi,  New J. Phys. {\bf 8} 235, (2006).
\bibitem{mohideen06}  F. CHen, U. Mohideen, G.L. Klimchitskaya and V. M. Mostepanenko, Phys. Rev. A {\bf 74} 022103 (2006).
  \bibitem{klich} O. Kenneth, I. Klich, A. Mann, and M. Revzen
Phys. Rev. Lett. {\bf 89}, 033001 (2002). 
  \bibitem{comment} D. Iannuzzi and F. Capasso
Phys. Rev. Lett. {\bf 91}, 029101 (2003). 
   \bibitem{born}   M. Born and E. Wolf, {\it Principles of Optics}, Pergamon Press (Oxford, 1980).
   \bibitem{maex}  M. H. Jo, H. H. Park, D. J. Kim,  S. H. Hyun, S. Y. Choi and J. T. Palik, J. Appl. Phys. {\bf 82} 1299 (1997). 
   \bibitem{sumiyoshi} T. Sumiyoshi, I. Adachi, T. Iijimat, R. Suda,  M. Yokoyama and H. Yokowaga, J. Non-Cryst. Solids {\bf 225}, 369-374 (1998).
\bibitem{hrubesh} L. W. Hrubesh, L. E. Keene ad V. R. Latorre, J. Mat. Res. {\bf 8}, 1736 (1993).
\bibitem{geis} S. Geiss, B. Muller and J. Fricke, J. Porous Mat. {\bf 7}, 423 (2000).  
\bibitem{aaa} C. Himcinschi, M. Friedrich, C. Murray, I. Streiter, S. E. Scchulz, T. Gessner and D. R. T. Zahn, Semicond. Sci. Technol. {\bf 16}, 806 (2001).
\bibitem{hrubesh}L. W. Hrubesh, L. E. Keene and V. R. Latorre, J. Mater. Res. {\bf 8}, 1736 (1993).
   \bibitem{looyenga} H. Looyenga, Physica {\bf 31}, 401 (1965).
\bibitem{palik}  E. D. Palik (Editor), {\it Handbook of Optical Properties of Solids}, (Academic Press, London 1985).
\bibitem{irinaau} I. Pyroshenko, A. Lambrecht and V. B. Svetovoy, New. J. Phys. {\bf 8}. 238 (2006). 
\bibitem{aucomment} The measurements are usually made between a $Au$ sphere and plate.
\bibitem{kochergin} V. Kochergin and H. Foll, Mat. Res. Soc. Symp. Proc., {\bf 881E}, CC3.1.1 (2005). 

 \end{references}
\end{document}